\documentclass[preprintnumbers,amsmath,11pt,amssymb,floatfix,superscriptaddress,showpacs,nofootinbib]{revtex4}
\usepackage{graphicx}
\usepackage{epsfig}
\usepackage{bm}
\usepackage{amsfonts}

\input epsf

\begin{document}

\title{\Large Holographic reconstruction of $f(G)$ Gravity for scale factors
pertaining to Emergent, Logamediate and Intermediate scenarios}

\author{Abdul Jawad}
\email{jawadab181@yahoo.com} \affiliation{Department of
Mathematics, University of the Punjab, Quaid-e-Azam Campus,
Lahore-54590, Pakistan.}

\author{Antonio Pasqua}
\email{toto.pasqua@gmail.com} \affiliation{Department of Physics,
University of Trieste, Trieste, Italy.}

\author{Surajit Chattopadhyay}
\email{surajit_2008@yahoo.co.in, surajcha@iucaa.ernet.in}
\affiliation{ Pailan College of Management and Technology, Bengal
Pailan Park, Kolkata-700 104, India.}

\date{\today}

\begin{abstract}

In this paper, we reconstruct the holographic dark energy in the
framework of $f(G)$ modified theory of gravity, where $G$ is
Gauss-Bonnet invariant. In this context, we choose the infrared
cut-off as Granda-Oliveros cut-off which is proportional to the
Hubble parameter $H$ and its first derivative with respect to the
cosmic time $t$. We reconstruct $f(G)$ model with the inclusion of
HDE and three well-known forms of the scale factor $a(t)$, i.e. the
emergent, the logamediate and the intermediate scale factors. The
reconstructed model as well as equation of state parameter are
discussed numerically with the help of graphical representation to
explore the accelerated expansion of the universe. Moreover, the
stability of the models incorporating all the scale factors is
checked through squared speed of sound $v_s^2$.

\end{abstract}


\maketitle

\section{Introduction}
It is established by cosmological observations (e.g. SNe Ia, CMB)
that the present universe is undergoing a phase of accelerated
expansion \cite{24,32}. The physical origin of this cosmic
acceleration has continued to be a deep mystery till date. Two
approaches have been adopted to explain this accelerated expansion.
They are:
\begin{enumerate}
    \item To introduce ``dark energy" (DE) in the right-hand side of the Einstein equation
in the framework of general relativity (recent reviews include
\cite{8,9,27,1});
    \item To introduce ``modified theory of gravity" by modifing the left-hand side of
the Einstein equation (recent reviews include \cite{22,18,21}).
\end{enumerate}
In an exhaustive review, Clifton et al. \cite{7} has thoroughly
discussed the motivations behind modified gravity theory. In Nojiri
and Odintsov \cite{19}, a detailed account of the advantages of the
modified gravity has been presented. According to Nojiri and
Odintsov \cite{19}, some significant advantages of modified gravity
theory are:
\begin{itemize}
  \item Modified gravity provides the very natural gravitational alternative for dark energy;
  \item Modified gravity presents very natural unification of the early-time inflation and late-time acceleration;
  \item Modified gravity serves as the basis for unified explanation of dark energy and dark matter;
  \item Modified gravity describes the transition from deceleration to acceleration in the universe evolution.
\end{itemize}
The development in the study of black hole theory and string theory
results the holographic principle \cite{4}. The holographic
principle ``calls into question not only the fundamental status of
field theory but the very notion of locality" \cite{4}. The
Holographic DE (HDE), based on the holographic principle, is one of
the most studied models of DE \cite{15, 11, 10, 12}. The present
work is devoted to the holographic reconstruction of one kind of
modified gravity theory, namely $f(G)$, via three choices to scale
factor pertaining to emergent, logamediate and intermediate
scenarios respectively. Thus, a three layered discussion would be
presented in the subsequent portion of this Section. Firstly, we
survey the choices of the scale factor. Choices of scale factors for
the scenarios mentioned above are:
\begin{enumerate}
    \item {\bf Emergent scenario} \cite{17}, where the
    scale factor has the form $a(t)=A\left(B+e^{n t}\right)^{\beta}$
    with $A>0,~B>0,~n>0,~\beta>1$.
    \item {\bf Intermediate scenario} \cite{2,3}, where $a(t)=\exp \left[B t^{\beta}\right]$ with
    $B>0,~0<\beta<1$.
    \item {\bf Logamediate scenario} \cite{2}, where $a(t)=\exp\left[B(\ln
    t)^{\beta}\right]$ with $B>0,~\beta>1$.\\
\end{enumerate}

Since reconstruction of a modified gravity is the basic purpose of
the present work, we now review the current status of research in
this direction. Reconstruction of DE has been addressed by a number
of researchers. Important references in this direction include
\cite{26,28,6,16}. Setare and Saridakis \cite{30} considered the
holographic and Gauss-Bonnet DE models separately to investigate the
conditions under which these models can be simultaneously valid.
This correspondence leads to accelerated expansion of the universe.
The same authors \cite{31} discussed the correspondence of HDE model
with canonical, phantom and quintom models minimally coupled to
gravity and observed consistent results about acceleration of the
universe. Chattopadhyay and Pasqua \cite{5} developed the
correspondence between $f(T)$ gravity and HDE model and discussed
the accelerated expansion of the universe. Jawad et al. \cite{13}
reconstructed the HDE model in context of $f(G)$ gravity with a
power-law solution. They found analytical solution for $f(G)$ model
in this scenario to discuss the EoS parameter, the stability
analysis as well as energy conditions to explore the current
expansion of the universe.

Motivated by these works, we have developed the reconstructed
scenario of HDE model with $f(G)$ model for three different scale
factors to represent a variety of models to discuss the accelerated
expansion of the universe. We have just extended our works
\cite{5,13}, reconstructing HDE model with GO cut-off in the frame
of $f(G)$ modified theory of gravity  for a variety of scale factors
(i.e. the emergent, the logamediate and the intermediate one). We
draw the numerically solution of the reconstructed scenario with the
help of three different forms of scale factors. The paper is
organized as follows: In the next section, we provide formalism of
$f(G)$ gravity and HDE model. Section \textbf{III} is devoted to the
brief discussion of scale factors. The reconstructed scenario via
scale factors is given in section \textbf{IV}. The last section
contains the concluding remarks.

\section{Brief overview of $f(G)$ gravity and HDE}

In this Section, we will describe some important features of $f(G)$
gravity and establish a correspondence between $f(G)$ and HDE. For
this purpose, we consider an action representing a special form of
$f(G)$ gravity model which contains an Einstein gravity term with
perfect fluid and an arbitrary function of Gauss-Bonnet term
\cite{25}. The action $S$ of this $f(G)$ modified gravity is given
by
\begin{equation}\label{1}
S=\int
d^4x\sqrt{-g}\left[\frac{1}{2\kappa^{2}}R+f(G)+\mathcal{L}_{m}\right],
\end{equation}
where
$G=R^{2}-4R_{\mu\nu}R^{\mu\nu}+R_{\mu\nu\lambda\sigma}R^{\mu\nu\lambda\sigma}$
(with $R$ representing the Ricci scalar curvature, $R_{\mu\nu}$ is
the Ricci curvature tensor and $R_{\mu\nu\lambda\sigma}$ denotes the
Riemann curvature tensor), $\kappa^{2}=8\pi G_N$ (with $G_N$ being
the gravitational constant), $g$ represents the determinant of the
metric tensor $g_{\mu \nu}$ and $\mathcal{L}_{m}$ represents the
Lagrangian of the matter present in the universe. The variation of
$S$ with respect to the metric tensor $g_{\mu\nu}$ generates the
field equations. For spatially flat FRW metric, the Ricci scalar
curvature $R$ and the Gauss-Bonnet invariant $G$, take the following
expressions respectively
\begin{equation}\label{2}
R=6(\dot{H}+2H^{2}), \quad G=24H^{2}(\dot{H}+H^{2}),
\end{equation}
where the upper dot represents the derivative with respect to the
cosmic time $t$.

The first FRW equation (with $8 \pi G_N=1)$ takes the form
\begin{equation}
H^{2}=\frac{1}{3}\left[Gf_{G}-f(G)-24\dot{G}H^{3}f_{GG}+\rho_{m}\right]=\frac{1}{3}\left(\rho_{G}+\rho_{m}\right),
\label{4}
\end{equation}
where $f_G$ and $f_{GG}$ represents, respectively, the first and the
second derivative of $f$ with respect to $G$, i.e.
$f_G=\frac{df}{dG}$ and $f_{GG}=\frac{d^2 f}{dG^2}$. Setare
\cite{29} has recently reconstructed $f(R)$ modified gravity from
HDE with IR cutoff as event horizon. Here we discuss a
reconstruction scheme of the above form of $f(G)$ gravity in HDE
framework taking Granda-Oliveros cut-off. The HDE density is defined
as \cite{29}
\begin{equation}
\rho_{\Lambda}=\frac{3}{L_{GO}^{2}}, \label{5}
\end{equation}
where $L_{GO}$ represents the Granda-Oliveros cut-off, which is
defined as
\begin{equation}
L_{GO} = \left(\lambda H^2 + \delta \dot{H}  \right)^{-\frac{1}{2}},
\label{6}
\end{equation}
where $\lambda$ and $\delta$ are constants. The dimensionless DE
density is obtained by dividing the energy density $\rho_{\Lambda}$
of DE with the critical energy density $\rho_{cr}=3H^{2}$, yields
\begin{equation}
\Omega_{\Lambda}=\frac{\rho_{\Lambda}}{\rho_{cr}}=\frac{c^{2}}{L_{GO}^{2}H^{2}}.
\label{7}
\end{equation}
Moreover, we define the EoS parameter $\omega_{eff}$ is defined as:
\begin{equation}
\omega_{eff}=\frac{p_{eff}}{\rho_{eff}},\label{8}
\end{equation}
where
\begin{eqnarray}\label{9}
\rho_{eff}&=&\rho + \frac{1}{2\kappa ^2}\left[  -f\left(G\right)
+24H^2\left( H^2 + \dot{H}
\right)f'\left(G\right)\right],\\\label{10} p_{eff}&=& p+
\frac{1}{2\kappa ^2}\left\{f\left(G\right)-24H^2\left( H^2 + \dot{H}
\right)f'\left(G\right)+ (24)8 H^{2}\left[ 6\dot{H}^{3} +6H\dot{H}
\ddot{H}+24H^{2}\dot{H}^{2}\right.\right.\\\nonumber &+&
\left.\left.6H^{3}\ddot{H}+8H^{4}\dot{H}+H^{2}\dddot{H}
\right]f''\left(G\right)+(24^2)8H^4\left( 2H^2
+H\ddot{H}+4H^2\dot{H} \right)^2f'''\left(G\right)  \right\}.
\end{eqnarray}
The prime denotes the derivative of the function with respect to
$G$.

\section{Discussion on the various scale factors}

Now we provide a glimpse on the main features of three scale factors
considered (i.e. the emergent, the logamediate and the intermediate
ones) and derive some related physical quantities from them to
discuss the reconstructed scenario.

\subsection{Emergent Scale Factor}
The emergent scale factor is given by \cite{17}
\begin{equation}\label{11}
a(t)=A\left[B+\exp(nt)\right]^{\beta},
\end{equation}
where $A,~B,~n$ and $\beta$ are positive constants. The avoid any
singularity, we have to take $B>0$ whereas $A$ should be greater
than zero for the positivity of scale factor. For $a<0$ or
$\beta<0$, the universe bears a big bang singularity, so for the
expanding model $a>0$ or $\beta>0$. This kind of scale factor was
recently used by Mukherjee et al. \cite{17} and Paul and Ghose
\cite{23}. The Hubble parameter and the Gauss-Bonnet invariant $G$,
taking into account the scale factor (\ref{11}), are given,
respectively, by
\begin{eqnarray} \label{12}
H = \frac{n \beta \exp (nt)}{B+ \exp (nt)},\quad G  = \frac{24n^4
\beta^3 \exp (3nt)\left[B+ \beta\exp (nt)\right]}{\left[B+ \exp
(nt)\right]^4}.
\end{eqnarray}

\subsection{Logamediate}

The next scale factor considered is the logamediate one, which has
the form \cite{2,3}
\begin{equation}\label{14}
a(t)=\exp\left[B(\ln t)^{\beta}\right],
\end{equation}
where $B\beta>0$ and $\beta >1$. Barrow and Nunes \cite{3} studied
the logamediate inflation, where the scale factor expands just the
logamediate scale factor expressed here. In this case, $H$ and $G$
take the form
\begin{eqnarray}\label{15}
H = \frac{B \beta \log \left( t \right)^{\beta -1}}{t}, \quad G =
\frac{24B^3\beta ^3 \log \left( t \right)^{3\beta -4}\left[
B\beta\log \left( t \right)^{\beta}   - \log \left( t \right) +\beta
-1     \right]} {t^4}.
\end{eqnarray}

\subsection{Intermediate}

The intermediate scale factor is given by \cite{2}
\begin{equation}\label{17}
a(t)=\exp\left[Bt^{\beta}\right],
\end{equation}
where $B>0$ and $0<\beta <1$. This kind of scale factor was used in
the work of Khatua and Debnath \cite{14}. In this case, the Hubble
parameter $H$ is given by
\begin{eqnarray}
H = B\beta t^{\beta -1}. \label{18}
\end{eqnarray}
Moreover, $G$ can be written as
\begin{eqnarray}
G = 24B^3\beta ^3 t^{3\beta -4}\left(  B\beta t^{\beta} + \beta -1
\right)  \label{19}
\end{eqnarray}

\section{Holographic reconstruction of $f(G)$ gravity}

In this Section, we discuss the HDE reconstruction of $f(G)$ gravity
with the help of three scale factors considered. In order to
incorporate the correspondence, we equate the energy density
$\rho_G$ of $f(G)$ gravity (given in Eq.(\ref{4})) and  energy
density $\rho_{\Lambda}$ of HDE model (given in Eq.(\ref{5})), which
leads to the following differential equation \cite{19,20}
\begin{eqnarray}
24H^3 \dot{G}^{-1} \ddot{f}(G) - (24 H^3 \dot{G}^{-2}\ddot{G} -
G\dot{G}^{-1}) \dot{f}(G) + f(G) = -3(\delta\dot{H} + \lambda H^2).
\label{20}
\end{eqnarray}
The search for analytical solution of this differential equation is
very difficult. Thus, we prefer the numerical solutions of this
equation for $f(G)$ to elaborate the behavior of EoS parameter and
squared speed of sound $v_s^2$ in reconstructed scenario of HDE in the
framework of $f(G)$ gravity.

\subsection{For Emergent Scale Factor}

We plot the $f(G)$ model of gravity obtained  numerically against the cosmic
$t$ for the emergent scale factor taking three different values of
$\beta$ (i.e. 1.2, 1.4 and 1.6) as shown in Figure 1. We
also use the following values for the remaining constants of the
scale factor: $A=2,~B=3.5,~n=4.5$ while $\lambda=1.5,~\delta=1$ for
the new HDE model. Initially, the graph shows a negatively
increasing behavior of $f(G)$ with respect to the cosmic time $t$.
Afterwards, it converges to zero as the time passes for all
considered values of $\beta$. Figure 2 represents the graph
of the squared speed of sound $v_s^2$ against the cosmic time $t$
considering the same values of constants given above. For this
quantity, we use the expression
$v_s^2=\frac{\dot{p}_{eff}}{\dot{\rho}_{eff}}$ along with
Eq. (\ref{20}). It exhibits the positive behavior for new $f(G)$ HDE
model for $\beta=1.2$ and $t>1.61$ corresponds to the stability of
the model. For $\beta=1.4$ and $1.6$, $v^2_s$ initially has positive
decreasing behavior leading to a stable model and it becomes
negative with the passage of time. Thus, it shows a classically
unstable model for the future epoch.

\begin{figure}[h]
\begin{minipage}{14pc}
\includegraphics[width=16pc]{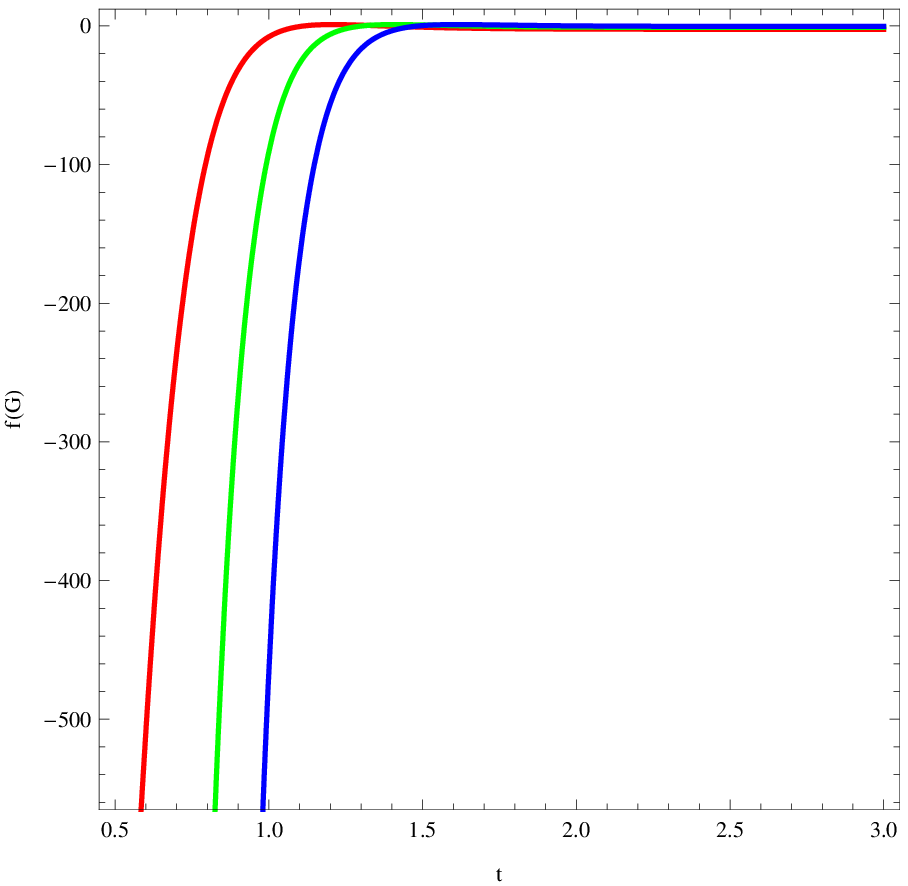}
\caption{\label{label}Plot of $f(G)$ versus $t$ for Emergent scale
factor with $\beta=1.2$ (red), $\beta=1.4$ (green) and $\beta=1.6$
(blue).}
\end{minipage}\hspace{3pc}%
\begin{minipage}{14pc}
\includegraphics[width=16pc]{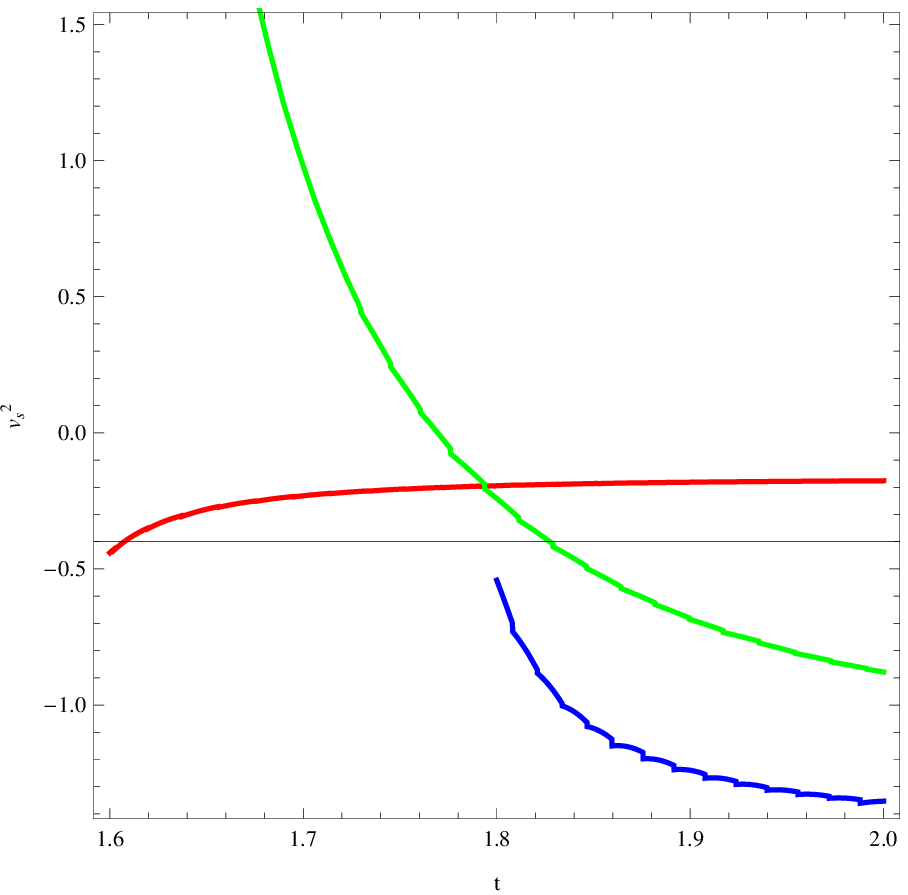}
\caption{\label{label}Plot of $v^2_s$ versus $t$ for Emergent scale
factor with $\beta=1.2$ (red), $\beta=1.4$ (green) and $\beta=1.6$
(blue).}
\end{minipage}\hspace{3pc}%
\end{figure}

The plot of the effective EoS parameter $\omega_{eff}$ against the
cosmic tine $t$ is given in Figure 3 with the help of
Eq.(\ref{20}). For $t=3.1$ and $\beta=1.2$, $\omega_{eff}$ crosses
the phantom divide line $\omega_{eff}=-1$ from phantom to
quintessence era of the expanding universe and becomes constant in
the quintessence dominated universe $(-1<\omega_{eff}\leq
-\frac{1}{3})$. It first expresses the matter dominated universe for
$\beta=1.6$ and then it experiences decreasing behavior with the
passage of time indicating the expanding universe with acceleration
in quintessence era. For $\beta=1.4$, ir represents the same era of
the universe. It is noted that the effective EoS parameter exhibits
small oscillations for increasing $t$, however it does not affect
the results corresponding to the parameter.
\begin{figure} \centering
\epsfig{file=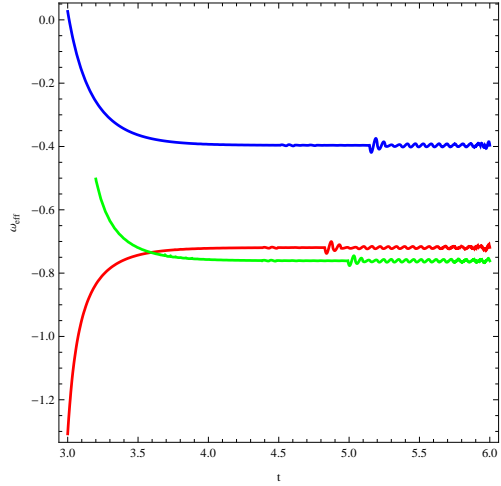,width=.40\linewidth} \caption{Plot of
$\omega_{eff}$ versus $t$ for Emergent scale factor with $\beta=1.2$
(red), $\beta=1.4$ (green) and $\beta=1.6$ (blue).}
\end{figure}

\subsection{For Logamediate Scale Factor}

We now consider the logamediate scale factor to discuss the
evolution trajectories of the new $f(G)$ HDE model, the squared
speed of sound $v_s^2$ and the effective EoS parameter
$\omega_{eff}$ taking into account Eq.(\ref{20}). Considering the
same values of $\lambda$ and $\delta$ for the emergent case with
$B=2$, we plot this scenario against cosmic time $t$ for three
different values of $\beta$, i.e. $\beta=1.001,~1.002,~1.003$. For
this purpose, we use Eqs.(\ref{15})  and (\ref{20}).
\begin{figure}[h]
\begin{minipage}{14pc}
\includegraphics[width=16pc]{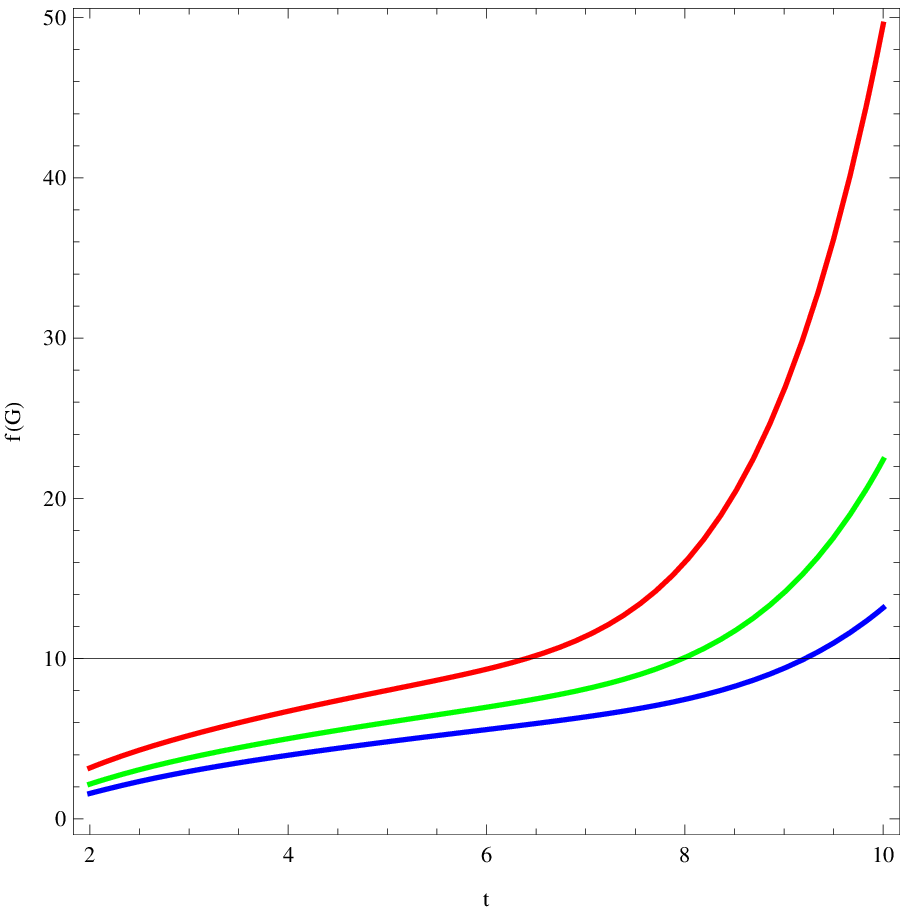}
\caption{\label{label}Plot of $f(G)$ versus $t$ for Logamediate
scale factor with $\beta=1.001$ (red), $\beta=1.002$ (green) and
$\beta=1.003$ (blue).}
\end{minipage}\hspace{3pc}%
\begin{minipage}{14pc}
\includegraphics[width=16pc]{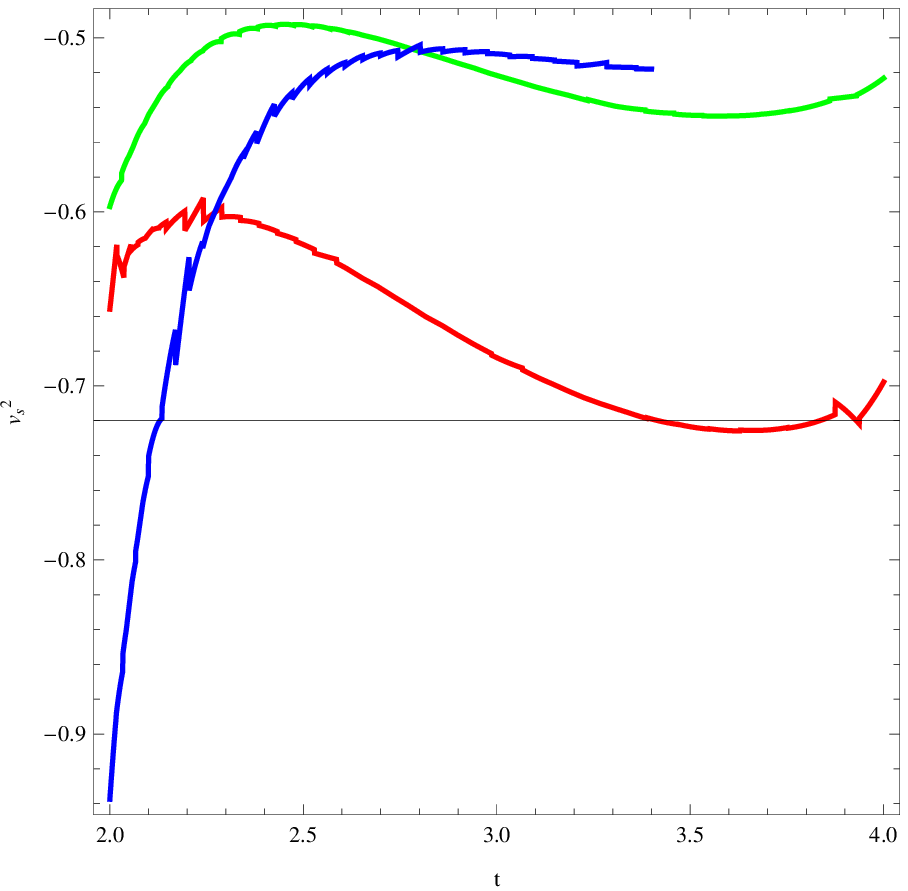}
\caption{\label{label}Plot of $v^2_s$ versus $t$ for Logamediate
scale factor with $\beta=1.001$ (red), $\beta=1.002$ (green) and
$\beta=1.003$ (blue).}
\end{minipage}\hspace{3pc}%
\end{figure}
\begin{figure} \centering
\epsfig{file=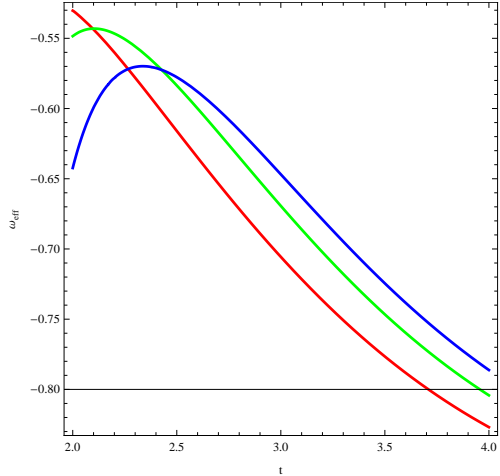,width=.40\linewidth} \caption{Plot of
$\omega_{eff}$ versus $t$ for Logamediate scale factor with
$\beta=1.001$ (red), $\beta=1.002$ (green) and $\beta=1.003$
(blue).}
\end{figure}
Figure 4 shows that the new $f(G)$ HDE model increases as
the cosmic time $t$ increases. It carries the more steeper plot for
the smaller values of $\beta$ and indicates the flatness for its
higher values. Thus the model represents the increasing and positive
behavior in the evolution of the universe. $v^2_s$ bears negative
behavior against the cosmic time $t$ as shown in Figure 5.
This implies instability of the model for all values of $\beta$
considered. Figure 6 represents the plot of $\omega_{eff}$ against
the cosmic time $t$ for all the choices of $\beta$ considered here.
It exhibits the quintessence region of the accelerated expansion of
the universe.

\subsection{For Intermediate Scale Factor}

By adopting the same procedure as for above scale factors, we
discuss the same quantities for intermediate scale factor. Figures
7, 8 and 9 describe, respectively, the
graphical behavior of new $f(G)$ HDE model, $v^2_s$ and
$\omega_{eff}$ against the cosmic time $t$ using the scale factor
considered here. We consider three different values of the parameter
$\beta$, i.e. $\beta=0.04,~0.05,~0.06$ along with $B=0.5$ and same
values of $\mu$ and $\lambda$ applied for the other scale factors.
The graph of new $f(G)$ HDE model displays the positive and
decreasing behavior with the passage of time for all values of
$\beta$. In the evolving universe, this model with intermediate
scale factor shows decaying behavior in the future epoch of the
universe. The plot of squared speed of sound against cosmic time $t$
demonstrates the stability of the model for all values of $\beta$.
However, the curve for $\beta=0.05$ may be examined as to come from
instability to stable and maintain it for the future epoch while the
other two curves stay always in stable region.
\begin{figure}[h]
\begin{minipage}{14pc}
\includegraphics[width=16pc]{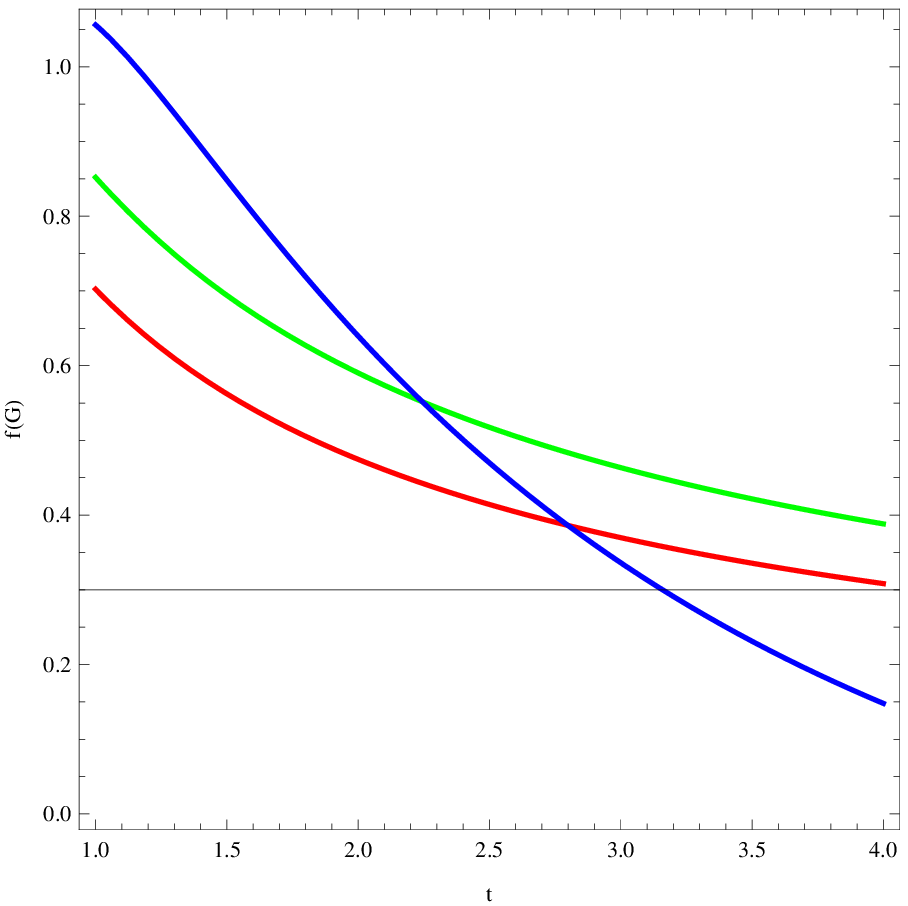}
\caption{\label{label}Plot of $f(G)$ versus $t$ for Intermediate
scale factor with $\beta=0.04$ (red), $\beta=0.05$ (green) and
$\beta=0.06$ (blue).}
\end{minipage}\hspace{3pc}%
\begin{minipage}{14pc}
\includegraphics[width=16pc]{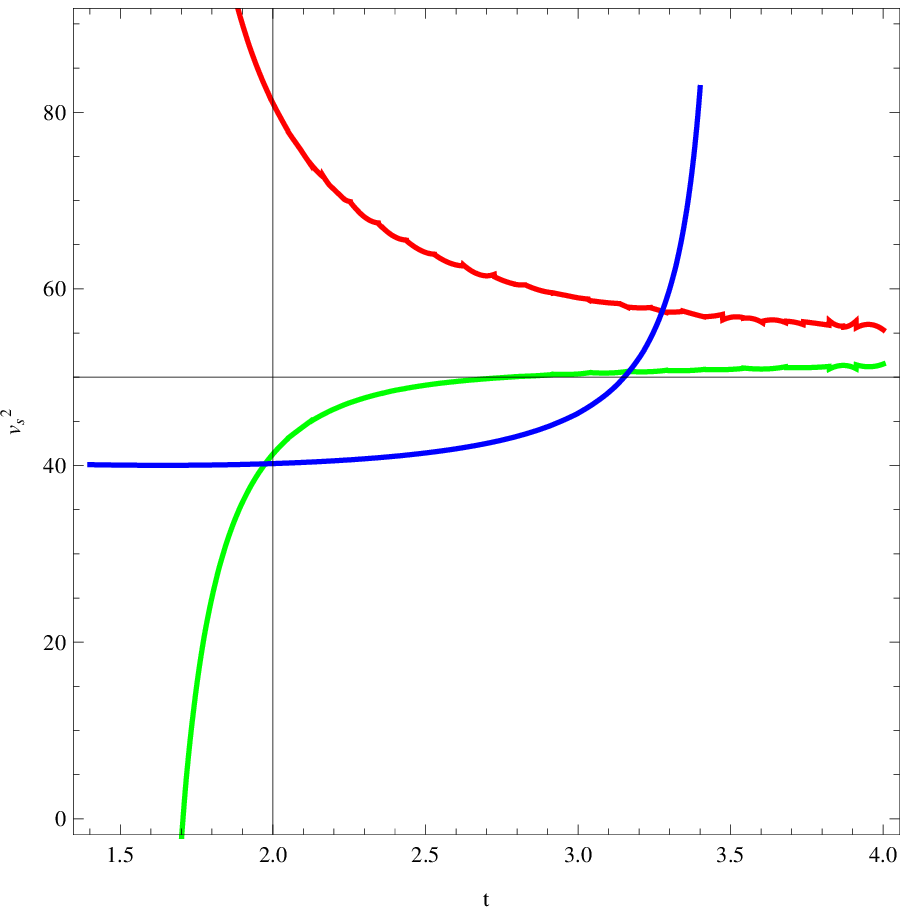}
\caption{\label{label}Plot of $v^2_s$ versus $t$ for Intermediate
scale factor with $\beta=0.04$ (red), $\beta=0.05$ (green) and
$\beta=0.06$ (blue).}
\end{minipage}\hspace{3pc}%
\end{figure}
\begin{figure} \centering
\epsfig{file=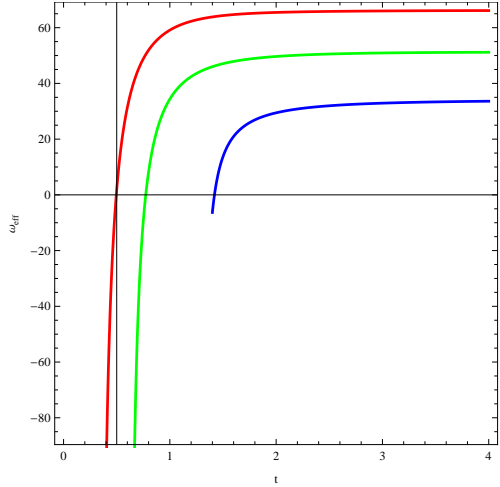,width=.40\linewidth} \caption{Plot of
$\omega_{eff}$ versus $t$ for Intermediate scale factor with
$\beta=0.04$ (red), $\beta=0.05$ (green) and $\beta=0.06$ (blue).}
\end{figure}

The EoS parameter bears negative behavior initially for a small
value of $t$ but it increases and becomes positive. It shows the
decelerating phase of the universe for the future epoch which is not
compatible with the recent observations accelerating universe.

\section{Concluding remarks}

The main aim of this work is an holographic reconstruction of $f(G)$
gravity taking as IR cut-off the recently proposed Granda-Oliveros cut-off, which is
proportional to the sum of Hubble parameter and its first time
derivative. We have taken here HDE with GO cut-off in order to check the
more viable/workable model than already discussed reconstruction
scenario. For this purpose, we have assumed three different models
of the scale factor $a(t)$, i.e. the emergent, the logamediate and
the intermediate one. The analytical solution of reconstructed model
is a hard job, thus we have preferred to obtain numerical solutions
with the help of graphs. We have constructed EoS parameter $\omega_{eff}$ as well
as an analysis about stability of the model with these scale
factors. The results of reconstructed scenario are as follows.

For the emergent scale factor, we have found that $f(G)$ converges
to zero after a negatively increasing behavior for all values of the
parameter $\beta$. Studying the squared speed of sound, it is
observed that the reconstructed model is classically unstable for
the future epoch for this kind of scale factor. The EoS parameter
shows different behaviors according to the values of $\beta$. For
$\beta = 1.2$, $\omega_{eff}$ is able to cross the phantom divide
line and it becomes constant in the quintessence dominated universe.

During the evolution of the universe, $f(G)$ has a positive and
increasing behavior for the logamediate scale factor. The squared
speed of the sound is always negative for all the values of the
parameter $\beta$, implying an instability of the model for this
kind of scale factor. The EoS parameter always stays in the
quintessence region of the accelerated expansion of the universe for
all values of the parameter $\beta$.

The graph of $f(G)$ has a positive and decreasing behavior as time
passes for all values of $\beta$ incorporating the intermediate
scale factor. The squared speed of the sound shows a general
stability of the model for this scale factor, even if the case with
$\beta = 0.05$ needs particular attention since it passes from the
instable to the stable region of the plot. The EoS parameter shows a
decelerating phase of the universe which appears to be not
compatible with recent cosmological observations.

\vspace{0.5cm}

{\bf Acknowledgment}

\vspace{0.5cm}

The first author (AJ) wishes to thank the Higher Education
Commission, Islamabad, Pakistan for its financial support through
the \textit{Indigenous Ph.D. 5000 Fellowship Program Batch-VII}.
The third author (SC) wishes to acknowledge the financial support
from Department of Science and Technology, Govt. of India under
Project Grant no. SR/FTP/PS-167/2011.

\end{document}